\documentclass[a4paper,twocolumn,english,aps,prl,showpacs,showkeys]{revtex4}
\usepackage[T1]{fontenc}
\usepackage[latin9]{inputenc}
\usepackage{amsmath}
\usepackage{graphicx}
\usepackage{amssymb}

\makeatletter
\usepackage{babel}
\makeatother

\begin{document}

\title{Upconversion cooling of Er-doped low-phonon fluorescent solids}

\author{Angel J. Garcia-Adeva, Rolindes Balda, and Joaquin Fernandez}

\affiliation{Departamento de Fisica Aplicada I, E.T.S. Ingenieria de Bilbao, Universidad
del Pais Vasco, Alda. Urquijo s/n, 48013 Bilbao, Spain}

\email{angel.garcia-adeva@ehu.es}



\begin{abstract}
We report on a novel mechanism for laser cooling of fluorescent solids
based on infrared-to-visible upconversion often found in rare-earth-doped
low-phonon materials. This type of optical cooling presents some advantages
with regards to conventional anti-Stokes cooling. Among them, it allows
to obtain cooling in a broader range of frequencies around the barycenter
of the infrared emitting band.
\end{abstract}
\maketitle
The demonstration by Epstein and coworkers \citep{EBE+95} of laser-induced
fluorescent cooling of an Yb-doped fluoride glass generated a surge
of interest in this phenomenon because it could provide the means
to fabricate the next-generation all-solid state compact optical cryocoolers
for aerospacial applications \citep{EBE98,EAE+99,RHR03,MM06} and
for the development of radiation-balanced lasers \citep{FGB06,Mun03}
that are amenable of multiple applications in the fields of optical
telecommunications and medicine. In this regard, it would be extremely
useful being able to use optical cryocoolers made of active materials
doped with different rare-earth ions that could be selected according
to the targeted application. Unfortunately, in spite of more than
two decades after the first experimental demonstration of this effect,
only a small number of ions could be used for this purpose. In particular,
anti-Stokes laser cooling has been demonstrated in a small number
of crystal and glass host materials \citep{RK07} doped with Yb \citep{EBE+95},
Tm \citep{HSE+00}, and Er \citep{FGB06}, due to the inherent characteristics
of the absorption and emission processes in rare earth ions. Two are
the main reasons that hinder the cooling efficiency: on the one hand,
the presence of nonradiative transitions between the energy levels
of the RE ion and, on the other hand, the presence of impurities in
the host matrix that gives rise to parasitic absorptions that generate
heat. Traditionally, host materials with very small phonons, so that
the quantum efficiency of the electronic levels involved in the fluorescent
transition is almost unity, with as few impurities as possible
have been used in order to overcome these two limitations. However,
even in the optimal case, the bulk cooling efficiency is still very
small \citep{FGB06,HSE+00}. In order to enhance the optical cooling
efficiency --most crucial for developing practical applications--
a number of ingenious innovations have been suggested, such as using
a multipass configuration, increasing the active medium length by
using optical fiber \citep{AB03}, or taking advantage of additional
cooling channels, such as cooling in the superradiance regime \citep{PS01},
energy transfer fluorescent cooling \citep{QHD+06}, or using nanocrystalline
powders doped with RE ions \citep{RK06}.

In this paper, we present a novel pathway to efficient anti-Stokes
laser cooling based on infrared-to-visible upconversion in RE-doped
host materials that we will term from now on \emph{upconversion cooling}.
This mechanism makes use of the efficient ir-to-vis upconversion phenomena
that is often found in RE-doped low-phonon host materials because
of the reduced multiphonon transition rates that makes possible for
the pump level to act as an intermediate photon reservoir from which
additional upconversion processes take place. This is the case, for
example, of Er-doped low-phonon potassium lead halide crystals KPb$_{2}X_{5}$,
($X=$Cl, Br) --such as the one we recently employed to demonstrate
anti-Stokes laser cooling in an Er-doped crystal--, where extremely
efficient ir-to-vis upconversion has been demonstrated \citep{GBF+05,BGV+04}.

\begin{figure}
\includegraphics[width=2.5in,keepaspectratio]{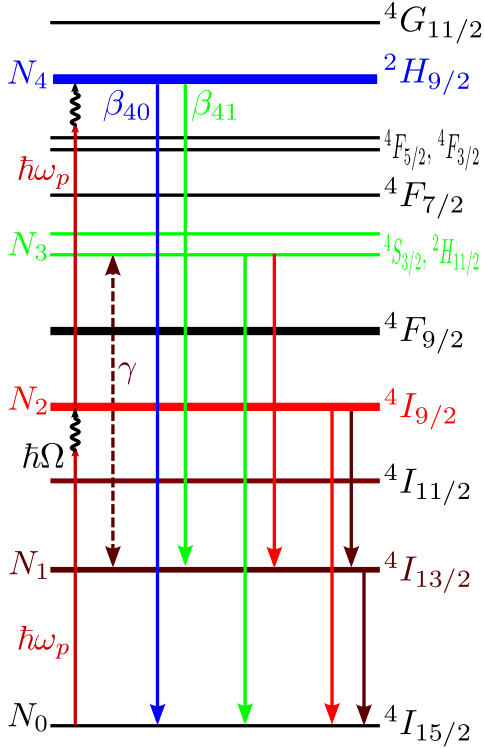}

\caption{\label{fig:levels}(color online). Experimental Er$^{3+}$ energy
levels in a KPb$_{2}$Cl$_{5}$ host matrix. The straight arrows indicate
the possible optical transitions that play a role in the electronic
dynamics. The wavy ones represent annihilation of phonons of energy
$\hbar\Omega$. The labels used in the rate equations are also indicated.}

\end{figure}
The upconversion cooling process is summarized in figure \ref{fig:levels}
by using the experimental energy levels of an Er-doped KPb$_{2}$Cl$_{5}$
crystal as a guiding example: upon cw pumping below the barycenter
of the $^{4}I_{9/2}$ level, an electronic population reservoir is
created in this level due to its long lifetime (2.4 ms). Part of this
population decays spontaneously to the $^{4}I_{15/2}$ level --emitting
infrared photons-- by means of a direct $^{4}I_{9/2}\rightarrow{}^{4}I_{15/2}$
transition or through a sequential $^{4}I_{9/2}\rightarrow{}^{4}I_{13/2}\rightarrow{}^{4}I_{15/2}$
one. In any case, an anti-Stokes cycle occurs in which the energy
of the fluorescent photons is slightly larger than the incident ones
and, thus, a small amount of thermal vibrational energy (temperature)
is removed from the system. Apart from this conventional laser cooling
mechanism, in this system there can also exist additional emissions
--in the visible range-- due to the presence of upconversion processes
\citep{BGV+04}. These can be of two types: on the one hand, there
is the possibility of sequential absorption by an electron in the
excited state of another pumping photon that promotes it to a higher
excited level from where it later decays spontaneously, the so called
excited state absorption (ESA) upconversion. On the other hand, there
can be an energy transfer upconversion (ETU) process in which two
electrons initially in the excited state of two different ions interact
with each other and, as a result, one of them decays to the ground
state whereas the other one is promoted to a higher excited state,
from where it later decays radiatively to the ground state. The population
dynamics of the electronic levels taking part in all these processes
can be cast in a very simple form by using a simple rate equation
formalism\begin{align}
N_{OAI} & =N_{0}+N_{1}+N_{2}+N_{3}+N_{4}\\
\frac{dN_{1}}{dt} & =-W_{10}\, N_{1}+\beta_{21}W_{21}\, N_{2}+\gamma\, N_{2}^{2}+\beta_{31}W_{31}\, N_{3}\nonumber \\
 & \qquad\qquad+\beta_{41}W_{41}\, N_{4}+\left(1-\eta_{e}^{(1)}\right)W_{1}^{rad}\, N_{1}\\
\frac{dN_{2}}{dt} & =\frac{P_{abs}^{r}}{h\nu_{p}}-\beta_{20}W_{20}\, N_{2}-\beta_{21}W_{21}\, N_{2}-2\gamma\, N_{2}^{2}\nonumber \\
 & -\sigma_{ESA}(\nu_{p})\frac{I_{p}(t)}{h\nu_{p}}\, N_{2}+\left(1-\eta_{e}^{(2)}\right)W_{2}^{rad}\, N_{2}\\
\frac{dN_{3}}{dt} & =-\beta_{30}W_{30}\, N_{3}-\beta_{31}W_{31}\, N_{3}+\gamma\, N_{2}^{2}\nonumber \\
 & +\left(1-\eta_{e}^{(3)}\right)W_{3}^{rad}\, N_{3}\\
\frac{dN_{4}}{dt} & =\sigma_{ESA}(\nu_{p})\frac{I_{p}(t)}{h\nu_{p}}\, N_{2}-\beta_{40}W_{40}\, N_{4}-\beta_{41}W_{41}\, N_{4}\nonumber \\
 & \qquad\qquad+\left(1-\eta_{e}^{(4)}\right)W_{4}^{rad}\, N_{4},\end{align}
where $N_{i}$ ($i=1\ldots5$) is the population of the $i$-th level
(see figure \ref{fig:levels} for the nomenclature of the levels),
$N_{OAI}$ is the total population of optically active ions (OAI)
in the sample, $W_{ij}=W_{i}^{rad}+W_{i\rightarrow j}^{nr}$ is decay
rate from level $i$ to level $j$ that includes both the radiative
and the non-radiative (phonon assisted) decay rates, $\beta_{i0(1)}$
is the branch ratio for the $i\rightarrow0(1)$ transition, $\gamma$
is the strength of the ETU process in which one electron in level
$^{4}I_{9/2}$ decays to the $^{4}I_{13/2}$ level and, at the same
time, another electron in a different ion is promoted from the $^{4}I_{9/2}$
level to the $^{4}S_{3/2}$ one, $\eta_{e}^{(i)}$ is the fraction
of photons emitted from the $i$-th level that actually leave the
sample (not reabsorbed), $\nu_{p}$ is the frequency of the pumping
laser, $P_{abs}^{r}$ is the power absorbed resonantly by the sample
at frequency $\nu_{p}$, $\sigma_{ESA}$ is the absorption cross-section
at that same frequency, and $I_{p}(t)$ is the intensity of the laser
beam. This set of equations cannot be solved analytically in the general
case. However, it is quite easy to calculate the steady-state populations
by using the condition $\frac{dN_{i}}{dt}=0$, ($i=0\ldots4$). Once
these populations are known, one can calculate the net
power deposited in the sample by using the balance equation\begin{multline}
P_{net}=P_{abs}^{r}+P_{abs}^{b}-\eta_{e}^{(2)}W_{2}^{rad}h\bar{\nu}_{2}N_{2}^{ss}\label{eq:pnet}\\
-\eta_{e}^{(3)}W_{3}^{rad}h\bar{\nu}_{3}N_{3}^{ss}-\eta_{e}^{(4)}W_{4}^{rad}h\bar{\nu}_{4}N_{4}^{ss},\end{multline}
where the frequencies $\bar{\nu}_{i}$ are the mean fluorescence frequency
of the $i$-th emitting band and $N_{i}^{ss}$ is the steady-state
population of the $i$-th level. In this expression, the first three
terms on the right hand side account for the standard anti-Stokes
absorption cooling mechanism, as described by Hoyt in Ref.~\onlinecite{hoytphd03}.
The fourth and fifth terms take into account the contribution from
the ESA and the ETU processes to cooling, respectively. The sample
temperature change can be easily calculated from \ref{eq:pnet} in
the low pump-depletion limit (a reasonable approximation in single
pass through the sample experiments) \citep{hoytphd03} and this quantity
can be cast in the form

\begin{multline}
\frac{\Delta T}{P_{in}}=\alpha_{b}+\alpha_{r}\left[1-\tilde{\eta}_{2}\frac{h\bar{\nu}_{2}}{h\nu_{p}}-\gamma\tilde{\eta}_{3}\left(\frac{\tilde{\eta}_{2}\tau_{2}^{rad}}{\eta_{e}^{(2)}}\right)^{2}\right.\\
\left.\times\alpha_{r}\frac{h\bar{\nu}_{3}}{h\nu_{p}}\frac{P_{in}}{h\nu_{p}}-\tilde{\eta}_{4}\frac{\tilde{\eta}_{2}\tau_{2}^{rad}}{\eta_{e}^{(2)}}\frac{\sigma_{ESA}(\nu_{p})}{A}\frac{h\bar{\nu}_{4}}{h\nu_{p}}\frac{P_{in}}{h\nu_{p}}\right],\label{eq:tempchange}\end{multline}
where $\alpha_{b}$ is a background absorption coefficient nearly
frequency independent \citep{hoytphd03}, $\alpha_{r}(\nu_{p})$ is
the resonant part of the absorption coefficient, $P_{in}$ is the
input laser power, and these three quantities are related to the power
absorbed by the sample (resonant or non-resonantly) through $P_{abs}^{r,nr}=P_{in}\frac{\alpha_{r,nr}}{\alpha_{T}}\left[1-\exp(-\alpha_{T}L)\right]$
with $\alpha_{T}=\alpha_{r}+\alpha_{nr}$ and $L$ the path length
of the laser beam in the sample, $\tilde{\eta}_{i}=\frac{\eta_{e}^{(i)}W_{i}^{rad}}{\eta_{e}^{(i)}W_{i}^{rad}+\beta_{i0}W_{i\rightarrow0}^{nr}+\beta_{i1}W_{i\rightarrow1}^{nr}}$
is the generalized quantum efficiency of the $i$-th level when one
takes into account the possibility of partial re-absorption of the
fluorescence, $\tau_{i}^{rad}=\frac{1}{W_{i}^{rad}}$ is the intrinsic
lifetime of the $i$-th level, and $A$ is the cross-sectional area
of the laser beam in the sample. By setting $\sigma_{ESA}=\gamma=0$,
one immediately recovers the standard model for anti-Stokes laser
cooling \citep{HSE+00}, in which the onset of cooling (the frequency
below which cooling occurs) is given by $\nu_{0}=\frac{\alpha_{r}}{\alpha_{T}}\bar{\nu}_{2}$.
However, the full model gives rise to a richer phenomenology. In order
to simplify the subsequent discussion, let us assume that the generalized
quantum efficiencies are almost unity and that is also the case for
the fraction of photons escaping the sample ($\tilde{\eta}_{i}=\eta_{e}^{(i)}\approx1$).
The first approximation is well justified in low-phonon materials
such as the potassium lead halides mentioned in the introduction for
which the multiphonon transition rates are almost negligible. This
is further confirmed by photothermal deflection spectroscopy measurements
in a number of RE-doped glasses and crystals. The second approximation
is more arguable, because it depends of a number of factors such as
the geometry of sample, its index of refraction, etc. However, with
regards to analyzing expression \ref{eq:tempchange}, we can always
make the substitution $\frac{\tau_{i}^{rad}}{\eta_{e}^{(i)}}\rightarrow\tilde{\tau}_{i}^{rad}$,
where the later lifetime is a renormalized one (that is, a photon
escaping fraction smaller than one can be interpreted as a longer
intrinsic lifetime of the emitting level), which does not alter any
of the following results whatsoever.

Let us start by considering a case in which standard anti-Stokes cooling
is negligible when compared with the upconversion channels, i.e.,
$\tau_{2}^{rad}\rightarrow\infty$. In this case, there is an input
power threshold for cooling, that is, we need to provide a minimum
input power to the sample in order to make $\Delta T<0$. This power
threshold is easily calculated to be\begin{equation}
P_{in}^{(0)}=\frac{\alpha_{T}}{\alpha_{r}}\frac{(h\nu_{p})^{2}}{\tau_{2}^{rad}}\left(h\nu_{up}\right)^{-1},\end{equation}
where \begin{equation}
h\,\nu_{up}=\gamma\tau_{2}^{rad}\alpha_{r}L\, h\bar{\nu}_{3}+\frac{\sigma_{ESA}(\nu_{p})}{A}h\bar{\nu}_{4}.\end{equation}
 Interestingly, this expression shows that upconversion cooling occurs
by means of any of the upconversion channels. This is quite useful
is systems where the operating upconversion mechanism can be selected
by adequately tuning the pumping frequency, such as it is the case
in potassium lead halide matrices \citep{GBF+05,BGV+04}. If we now
go back to the general model, another important conclusion coming
out from equation \ref{eq:tempchange} is that the onset of cooling,
that is, the pumping frequency below which cooling occurs is larger
when upconversion is present. The exact cutoff frequency depends on
both $\alpha_{r}(\nu_{p})$ and $\sigma_{ESA}(\nu_{p})$ and, as such,
cannot be put in a closed form in the general case. However, near
the resonance, where these two functions can be approximated by their
values at the maximum --$\alpha_{0}$ and $\sigma_{0}$-- and the
constant background absorption can be neglected, the cutoff frequency
can be put in the form\begin{equation}
\nu_{0}^{up}\approx\nu_{0}+\frac{\nu_{up}\,\tau_{2}^{rad}\, P_{in}}{{h\nu}_{0}}.\end{equation}
It is important to notice that $\nu_{0}^{up}>\nu_{0}$. This, in fact,
is quite reasonable, as one has additional channels for extracting
energy from the system when upconversion is present, so that cooling
can be more efficient. What it is really interesting is the fact that
if the pumping beam is very intense or the upconversion process very
efficient ($\nu_{up}\gg1$), the onset of cooling could be located
above the barycenter of the pumping level, that is, in its Stokes
part.

In conclusion, we have presented a novel mechanism for laser cooling
based on infrared-to-visible upconversion processes. It has been analyzed
by using a simple rate equation formalism from which it is concluded
that upconversion cooling presents a number of advantages with regards
to standard anti-Stokes cooling. Among these there is the possibility
to control the onset of cooling by using the pumping power or to select
the active upconversion mechanism by adequately tuning the pumping
frequency, which can be very useful for frequency selective applications.
We hope that the results presented in this work will encourage other
researchers to further investigate the role that upconversion cooling
could play in the development of highly efficient solid state refrigeration
applications.

\bibliographystyle{apsrev}

\end{document}